\documentclass[12pt]{article}
\usepackage{amstex}
\usepackage{amssymb}
\usepackage{babel}
\usepackage{latexsym}
\begin{document}
\newfont{\fraktgm}{eurm10 scaled 1728}
\newcommand{\graktur}{\baselineskip12.5pt \fraktgm}
\newfont{\fraktfm}{eurm10 scaled 1440}
\newcommand{\frakture}{\baselineskip12.5pt\fraktfm}
\newfont{\fraktrm}{eurm10}
\newcommand{\fraktur}{\baselineskip12.5pt\fraktrm}
\newfont{\fraktem}{eurm6}
\newcommand{\fraktr}{\baselineskip12.5pt\fraktem}
\protect\newtheorem{principle}{Principle}
\protect\newtheorem{theo}{Theorem}
\protect\newtheorem{prop}{Proposition}
\protect\newtheorem{lem}{Lemma}
\protect\newtheorem{co}{Corollary}
\protect\newtheorem{de}{Definition}
\begin{titlepage}
%\centerline{\normalsize DESY 96 - 176 \hfill ISSN 0418 - 9833} 
%\centerline{\normalsize August 1996 \hfill} 
\centerline{\normalsize gr-qc/9608066 v4 \hfill} 
\vskip.6in 
\begin{center} 
{\graktur On the Classification of Decoherence Functionals} 
\vskip.6in 
{{\Large \frakture Oliver Rudolph} $^*$} 
\vskip.3in 
{\normalsize \sf II.~Institut f\"ur Theoretische Physik, 
Universit\"at Hamburg} 
\vskip.05in 
{\normalsize \sf Luruper Chaussee 149} 
\vskip.05in 
{\normalsize \sf D-22761 Hamburg, Germany}
\vskip.7in
\end{center}
\normalsize
\vfill
\begin{center}
{ABSTRACT}
\end{center}
\smallskip
\noindent The basic ingredients of the \emph{consistent histories 
approach to quantum mechanics} are the space of histories 
and the space of decoherence 
functionals. In this work we extend the classification theorem for 
decoherence functionals proven by Isham, Linden and Schreckenberg 
to the case where the space of histories is the lattice of 
projection operators on an arbitrary separable or non-separable 
complex Hilbert space of dimension greater than two. \\ 
\\ \bigskip \noindent
\centerline{\vrule height0.25pt depth0.25pt width4cm \hfill}
\noindent
{\footnotesize $^*$ Internet: rudolph@@x4u2.desy.de} 
\end{titlepage}
\newpage
\section{Introduction}
\indent The consistent histories approach to quantum mechanics has 
attracted much interest in the last years. The consistent 
histories approach has enriched and deepened our understanding of 
nonrelativistic quantum mechanics and 
in particular of the interpretation of standard Hilbert space 
quantum mechanics. There is also hope that the consistent 
histories approach may be a guide towards the construction of 
history theories generalizing standard Hilbert space quantum 
mechanics. This hope is supported by the observation that general 
quantum history theories exhibit a much richer structure than 
standard quantum mechanics \cite{Isham94,Isham96}. \\ 

The consistent histories approach to nonrelativistic quantum 
mechanics has been inaugurated in a seminal paper by Griffiths 
\cite{Griffiths84} and further developed by Griffiths 
\cite{Griffiths96}, by Omn\`{e}s \cite{Omnes88a}-\cite{Omnes94}, 
by Isham \cite{Isham94,Isham96}, by Isham and Linden 
\cite{IshamL94}, by Isham, Linden and Schreckenberg 
\cite{IshamLS94}, by \linebreak[3] Schreckenberg 
\cite{Schreckenberg95}, by 
Pulmannov\'a \cite{Pulmannova95} and by this author 
\cite{Rudolph96}. In a series of interesting publications 
Gell-Mann and Hartle \cite{GellMann90a} have studied quantum 
cosmology and the path integral approach to relativistic quantum 
field theory in the framework of consistent histories. Further 
important developments and a critical examination of the 
consistent histories approach can be found in the work by Dowker 
and Kent \cite{Dowker96} and Kent \cite{Kent96}. \\ 

The consistent histories approach asserts 
that quantum mechanics provides a realistic description of 
\emph{individual} quantum mechanical systems, regardless of 
whether they 
are open or closed and regardless of whether they are observed or 
not. Probabilities are thought of as measures of 
propensities. To avoid confusion it should be stressed that the 
term \emph{realistic description} is not meant here in the sense 
of determinism or hidden variable theories. \\ 
The basic ingredients in the consistent histories approach are the 
space of histories on the one hand and the space of decoherence 
functionals on the other hand. The 
histories are identified with the general possibilities or 
properties of a quantum system. In a somewhat different language 
histories may be said to represent \emph{temporal events} or 
simply \emph{events}. The 
probabilities associated with histories are interpreted as 
measures of the tendency that certain 
\emph{histories} will be realized in a single system. The 
assignment of probabilities to certain histories is only 
admissible when these histories belong to a common Boolean lattice  
of histories which satisfies some consistency condition 
\cite{Isham94,Griffiths84,Griffiths96,Omnes88a,Omnes94,Rudolph96}. 
\\ 

In standard Hilbert space quantum mechanics the state of some 
quantum mechanical system comprises all probabilistic 
predictions of quantum mechanics for the system in question. 
This idea of the notion of state can be carried over to general 
quantum history 
theories: it is in this sense that decoherence functionals can be 
said to represent the \emph{state} of a system described by a 
quantum history theory. \\ 

To get insight into the possible structure of general history 
theories it is worthwhile to study the (algebraic) structure of 
the space of decoherence functionals for general quantum history 
theories in some detail. 
In particular --- as also stressed by Isham, Linden and 
Schreckenberg \cite{IshamLS94} --- the classification of 
decoherence functionals is an important problem. It is equivalent 
to the classification of states in quantum history theories. As is 
well-known, the analogous problem in standard quantum mechanics 
has been completely solved 
by Gleason \cite{Gleason57,Dvurecenskij93}. \\ \\ 
This work is organized as follows: in Section 2 the classification 
theorem for decoherence functionals is formulated and proved. 
Section 2.1 is devoted to an exposition of some necessary 
basic definitions and propositions. In 
Section 2.2 we give an alternative proof for the 
classification theorem for decoherence functionals in the case 
when the set of histories is the set of projection operators on a 
finite-dimensional complex Hilbert space. This 
classification theorem has first been proven by Isham, Linden and 
Schreckenberg \cite{IshamLS94}. Our proof is based on methods used 
by Cooke, Keane and Moran in their elementary proof of Gleason's 
theorem \cite{CookeKM85} and differs from the proof given by 
Isham, Linden and Schreckenberg in that we do not use Gleason's 
theorem directly. However, our proof makes use of a theorem due to 
Wright which is in turn based on the solution of the 
Mackey-Gleason problem \cite{Wright95,BunceW94}. 
In Section 2.3 we consider the case that 
the space of histories is the set  $\mathcal{P}(\mathfrak{H})$ of 
projection operators on some infinite-dimensional separable or 
non-separable complex Hilbert space $\mathfrak{H}$ 
and extend the classification theorem for decoherence functionals 
to this case. It is perhaps worthwhile to mention 
that our result is also valid if we identify the set of histories 
with the set of effects $\mathfrak{E}(\mathfrak{H})$ on some 
Hilbert space (as done in \cite{Rudolph96}) since every 
ultraweakly continuous normal decoherence functional on 
$\mathcal{P}(\mathfrak{H})$ can be 
uniquely extended to a functional on $\mathfrak{E}(\mathfrak{H})$, 
as shown in Corollary \ref{L1} in Section 2.4. Section 3 
presents our summary. \\

Throughout this work we will make use of Dirac's well-known ket 
and bra notation to denote vectors in Hilbert space and dual 
vectors in the dual Hilbert space respectively. 

\section{The Classification Theorem}
\subsection{Preliminaries}
Consider a history theory where the set of histories can be 
identified with the set $\mathcal{P}(\mathfrak{H})$ of projection 
operators on some separable or non-separable complex 
Hilbert space $\mathfrak{H}$. 
A functional $d : \mathcal{P}({\mathfrak H})\times 
\mathcal{P}({\mathfrak H})\rightarrow {\mathbb{C}}, 
(h,k)\longmapsto d(h,k)$ will 
be called a \textsc{decoherence functional on} 
$\mathcal{P}(\mathfrak{H})$ if the following conditions are 
satisfied for all $h,h^{\prime}, k\in \mathcal{P}({\mathfrak 
H}):$ \begin{itemize} \item $d(h,h)\in {\mathbb{R}},$ and 
$d(h,h)\geq 0.$ \item $d(h,k)=d(k,h)^{*}.$ \item $d(1,1)=1$ and 
$d(0,h)=0.$ \item $d(h\vee h^{\prime}, k) = d(h,k) + 
d(h^{\prime},k),$ whenever $h\perp h^{\prime}.$ \end{itemize}
By definition every decoherence functional is finitely 
additive in both arguments. We say that a decoherence functional 
$d$ is $\sigma$-\textsc{additive} in both arguments whenever
\begin{itemize} \item $d \left( \bigvee_{i \in \mathcal{I}} h_i, 
k \right) = \sum_{i \in \mathcal{I}} d(h_i,k),$ \end{itemize} 
whenever $\{ h_i \}_{i \in \mathcal{I}}$ is a countable set of 
mutually orthogonal histories and $k$ is an arbitrary history. We 
say that a decoherence functional $d$ is 
$\sigma$-\textsc{summable} in both arguments if $d$ is 
$\sigma$-additive and $\sum_{i \in \mathcal{I}} d(h_i,k)$ 
converges absolutely for all $k$ and all 
countable families $\{ h_i \}_{i \in \mathcal{I}}$ of mutually 
orthogonal histories. Moreover, we say that a decoherence 
functional $d$ is \textsc{completely additive} in both arguments 
if for any set $ \{ h_j \}_{j \in \mathcal{J}}$ of mutually 
orthogonal histories and for all $k$ the family $\{ d(h_i,k) \}$ 
is summable and \begin{itemize} \item $d \left( \bigvee_{j \in 
\mathcal{J}} h_j, k \right) = \sum_{j \in \mathcal{J}} d(h_j,k).$ 
\end{itemize}  

Every decoherence functional $d$ can be used to define 
\emph{consistent Boolean sublattices} of $\mathcal{P}({\mathfrak 
H})$ such that $d$ induces a probability measure on these 
consistent Boolean lattices. Clearly, a probability measure 
induced by a decoherence functional $d$ on a consistent Boolean 
lattice $\mathcal{B}_c$ is $\sigma$-additive if and only if $d$ is 
$\sigma$-additive in both arguments on $\mathcal{B}_c$ and is 
completely additive if and only if $d$ is completely additive in 
both arguments on $\mathcal{B}_c$. \\ 

Throughout this work $\mathcal{B}(\mathfrak{H})$ denotes the set 
of all bounded operators on a complex Hilbert space $\mathfrak{H}$ 
and $\mathcal{S}(\mathfrak{H})$ denotes the set of all unit 
vectors in the complex Hilbert space $\mathfrak{H}$. Moreover, 
$\mathfrak{E}(\mathfrak{H})$ denotes the set of all effect 
operators on $\mathfrak{H}$, i.e., the set of all Hermitean 
operators $E$ on $\mathfrak{H}$ with $0\leq E \leq 1$. The set 
$\mathfrak{E}(\mathfrak{H})$ carries the structure of a D-poset 
\cite{Kopka94}. In \cite{Rudolph96} it has been shown that 
$\mathfrak{E}(\mathfrak{H})$ can be supplied with countably many 
different D-poset structures. However, all the D-poset structures 
on $\mathfrak{E}(\mathfrak{H})$ considered in \cite{Rudolph96} are 
isomorphic and thus it is enough to consider the \emph{canonical} 
D-poset structure on $\mathfrak{E}(\mathfrak{H})$. We recall that 
the canonical D-poset structure on $\mathfrak{E}(\mathfrak{H})$ is 
given by a partially defined addition $\oplus$ on 
$\mathfrak{E}(\mathfrak{H})$: for $e_1, e_2 \in 
\mathfrak{E}(\mathfrak{H})$ the 
expression $e_1 \oplus e_2$ is defined if (and only if) $e_1 + e_2 
\leq 1$ by $e_1 \oplus e_2 := e_1 + e_2$. \\

If $\mathfrak{T}$ is a topology on 
$\mathcal{P}(\mathfrak{H})$, then a decoherence functional $d$ is 
called \textsc{bi-continuous with respect to the topology} 
$\mathfrak{T}$ if $d$ is continuous in both arguments with respect 
to the topology $\mathfrak{T}$. In the present work we use the 
standard nomenclature for topologies on 
$\mathcal{B}(\mathfrak{H})$, see, e.g., \cite{Dvurecenskij93}. 
We say that a decoherence functional $d$ 
is \textsf{ultraweakly bi-continuous} if $d$ is continuous in both 
arguments with respect to the ultraweak operator topology on 
$\mathcal{P}(\mathfrak{H})$. Every ultraweakly bi-continuous 
decoherence functional is obviously also continuous with respect 
to every stronger topology but not vice versa. The results in 
Section 2.3 are formulated for ultraweakly bi-continuous 
decoherence functionals. Since for norm bounded sequences of 
operators the notions of weak and ultraweak convergence coincide, 
it is clear that the results in Section 2.3 below are also valid 
for weakly bi-continuous decoherence functionals. However, the 
classification theorem for decoherence functionals on 
infinite-dimensional Hilbert spaces is 
in general not valid for decoherence 
functionals which are bi-continuous with respect to a stronger 
topology than the ultraweak topology. \\ \\
\noindent \textbf{Remark 1} Let $\mathfrak{H}$ denote a 
complex Hilbert space with $\dim(\mathfrak{H}) >2$. Wright 
\cite{Wright95,BunceW94} has proven the important 
general result that a bounded decoherence functional $d$ on 
$\mathcal{P}(\mathfrak{H})$ can be uniquely extended to a 
bilinear bounded functional $\mathcal{D}$ on 
$\mathcal{B}(\mathfrak{H})$. The extension $\mathcal{D}$ is 
continuous in both arguments with respect to the norm topology on 
$\mathcal{B}(\mathfrak{H})$ and thus every bounded decoherence 
functional $d$ is necessarily bi-continuous with respect to the 
topology induced on $\mathcal{P}(\mathfrak{H})$ by the norm on 
$\mathcal{B}(\mathfrak{H})$. \\ \\
Let $\mathfrak{H}$ denote a complex Hilbert space. We 
denote by $\mathfrak{H}_0$ the everywhere dense subset 
of $\mathfrak{H} \otimes \mathfrak{H}$ generated by the simple 
vectors of the form $\vert \phi \rangle \otimes \vert \psi 
\rangle$, where $\vert \phi \rangle, \vert \psi \rangle \in 
\mathfrak{H}$. That is, $\mathfrak{H}_0$ contains all finite 
linear 
combinations of simple vectors. For all $\vert \Psi_0 \rangle \in 
\mathfrak{H}_0$ we denote the one-dimensional projection operator 
onto $\vert \Psi_0 \rangle$ by $P_{\Psi_0}$. We denote the set of 
all such projection operators by $\mathcal{P}(\mathfrak{H}_0)$. 
Moreover, we denote 
the set of all projection operators in $\mathcal{P}(\mathfrak{H} 
\otimes \mathfrak{H})$ which can be written as a finite sum 
$\sum_j P_{\Psi_j}$ of pairwise orthogonal projection operators 
$P_{\Psi_j} \in \mathcal{P}(\mathfrak{H}_0)$ by 
$\mathcal{P}_{fin}(\mathfrak{H}_0)$. The set of all projection 
operators in $\mathcal{P}(\mathfrak{H} \otimes \mathfrak{H})$ 
which can be written as a ultraweakly converging sum $\sum_j 
P_{\Psi_j}$ of pairwise orthogonal projection operators 
$P_{\Psi_j} \in \mathcal{P}(\mathfrak{H}_0)$ will be denoted by 
$\mathcal{P}_{\infty}(\mathfrak{H}_0)$. If $\mathfrak{H}$ is 
finite-dimensional, then obviously $\mathfrak{H}_0 = 
\mathfrak{H} \otimes \mathfrak{H}$. 
\begin{prop} Let $\mathfrak{H}$ be a complex Hilbert space 
with $\dim(\mathfrak{H}) > 2$
and let $d$ denote a bounded decoherence functional on 
$\mathcal{P}({\mathfrak H})$, then $d$ 
can be uniquely extended to a 
functional $\widehat{d} : \mathcal{P}_{fin}(\mathfrak{H}_0) \to 
\mathbb{C}$ satisfying $\widehat{d}(h \otimes k) = d(h,k)$ for all 
$h,k \in \mathcal{P}(\mathfrak{H})$. Moreover, $\widehat{d}$ is 
additive for orthogonal projection operators, i.e., 
$\widehat{d}(P_1 + P_2) = \widehat{d}(P_1) + \widehat{d}(P_2)$ for 
all $P_1, P_2 \in \mathcal{P}_{fin}(\mathfrak{H}_0)$ with $P_1 
\perp P_2$. \label{L6} If $\mathfrak{H}$ is finite-dimensional, 
then $\widehat{d}$ is bounded. \end{prop}
\begin{lem} Let $E_1, ..., E_{n+m}, F_1, ..., F_{m+n} \in 
\mathcal{B}(\mathfrak{H})$, 
then  $(\sum_{i=1}^n E_i \otimes F_i) + (\sum_{i=n+1}^{n+m} E_i 
\otimes F_i) =0$ if and only if there is an $(n+m) \times (n+m)$ 
complex matrix $[c_{ik}]$ such that
\begin{eqnarray*} \sum_{i=1}^{n+m} c_{ik} E_i & = & 0, (k=1, ..., 
n+m), \\ \sum_{k=1}^{n+m} c_{ik} F_k & = & F_i, (i=1, ..., n+m). 
\end{eqnarray*} \label{L4} \end{lem} 
The assertion of Lemma \ref{L4} is exactly Proposition 
11.1.8 (i) in \cite{KadisonR86}. \\ \\ \noindent \textbf{Proof of 
Proposition \ref{L6}:} For simple projection operators 
of the form $h \otimes k$ with $h,k \in \mathcal{P}(\mathfrak{H})$ 
we define $\widehat{d}(h\otimes k) := 
d(h,k)$. We denote by $\mathfrak{H}_0$ the everywhere dense subset 
of $\mathfrak{H} \otimes \mathfrak{H}$ generated by the simple 
vectors of the form $\vert \phi \rangle \otimes \vert \psi 
\rangle$. That is, $\mathfrak{H}_0$ contains all finite linear 
combinations of simple vectors. Let $\vert \Psi_0 \rangle \in 
\mathcal{S}(\mathfrak{H}_0) := \mathcal{S}(\mathfrak{H} \otimes 
\mathfrak{H}) \cap \mathfrak{H}_0$, then $\vert \Psi_0 \rangle$ 
can be 
written as $\vert \Psi_0 \rangle = \sum_{j=1}^N \kappa_j \vert 
\phi_j \rangle \otimes \vert \psi_j \rangle$, with $\vert \phi_j 
\rangle, \vert \psi_j \rangle \in \mathcal{S}(\mathfrak{H})$ for 
all $j$. Denote the projection operator on 
$\vert \Psi_0 \rangle$ by $P_{\Psi_0} = \vert \Psi_0 \rangle 
\langle \Psi_0 \vert$ and define $\widehat{d}(P_{\Psi_0}) := 
\sum_{i=1}^N \sum_{j=1}^N \kappa_i \kappa_j^* \mathcal{D} \left( 
\vert \phi_i \rangle \langle \phi_j \vert, \vert \psi_i \rangle 
\langle \psi_j \vert \right)$ 
where $\mathcal{D}$ denotes the unique extension of $d$ to 
$\mathcal{B}(\mathfrak{H})$ mentioned in Remark 1. Lemma \ref{L4} 
implies that the such defined $\widehat{d}(P_{\Psi_0})$ is 
independent of the particular representation of $\vert \Psi_0 
\rangle \in \mathcal{S}(\mathfrak{H}_0)$ chosen (see Proposition 
11.1.8 (ii) in 
\cite{KadisonR86}). Thus we have extended $d$ to a functional 
on the set of projection operators of the form $P_{\Psi_0}$. 
Now let $P_M \in \mathcal{P}_{fin}(\mathfrak{H}_0)$. Then by 
definition $P_M$ can be written as a finite sum $P_M = 
\sum_{j=1}^M P_{\Psi_j}$ of mutually orthogonal projection 
operators $P_{\Psi_j} \in \mathcal{P}(\mathfrak{H}_0)$. We define 
$\widehat{d}(P_M) := \sum_{j=1}^M \widehat{d}(P_{\Psi_j})$. Again 
Lemma \ref{L4} implies that $\widehat{d}(P_M)$ is well-defined and 
independent of the particular decomposition of $P_M$ chosen. 
If $\mathfrak{H}$ is $n$-dimensional, then $\widehat{d}$ is 
bounded by $n^4 K$, where $K$ is a bound of $\mathcal{D}$. 
\hfill $\blacksquare$ \\ \\
We say that a bounded decoherence functional $d$ is 
\textsc{proper} if the unique extension $\widehat{d} : 
\mathcal{P}_{fin}(\mathfrak{H}_0) \to 
\mathbb{C}$ from Proposition \ref{L6} is bounded. Let $P \in 
\mathcal{P}_{\infty}(\mathfrak{H}_0)$ and $P = \sum_{j \in J} p_j$ 
one of its representations with $p_j \in 
\mathcal{P}(\mathfrak{H}_0)$. If $d$ is proper, then the net 
$\big\{ \widehat{d} \left(\sum_{j \in J_0} p_j \right) \vert J_0 
\subset J, J_0$ finite $\big\}$ \linebreak[3] possesses a 
converging subsequence. We say that a bounded 
decoherence functional $d$ is \textsc{normal} if $\widehat{d}$ can 
be unambiguously extended to a completely additive functional 
$\widetilde{d}: \mathcal{P}_{\infty}(\mathfrak{H}_0) \to 
\mathbb{C}$. \\ 

We introduce some notations and terminology: 
A map $m : \mathcal{P}(\mathfrak{H}) \to \mathbb{R}$ such that 
\begin{eqnarray} m(0) & = & 
0, \\ m \left( \bigvee_{i \in \mathcal{I}} p_i \right) & = & 
\sum_{i \in \mathcal{I}} m \left( p_i \right), \label{E0} 
\end{eqnarray}
whenever $\{ p_i \}_{i \in \mathcal{I}}$ is a system of mutually 
orthogonal projection operators in $\mathcal{P}(\mathfrak{H})$ is 
said to be (i) a \textsc{finitely additive signed measure}, (ii) a 
\textsc{signed measure}, or (iii) a \textsc{completely additive 
signed measure} if Equation \ref{E0} holds for every (i) finite, 
(ii) countable, or (iii) arbitrary index set $\mathcal{I}$ 
respectively. 
A finitely additive signed measure is said to be \textsc{Jordan} 
if it can be written as a difference of two positive finitely 
additive measures. \\ 

A map 
$f : \mathcal{S}(\mathfrak{H}) \to \mathbb{R}$ is called a 
\textsc{frame function} if there is a constant $\omega \in 
\mathbb{R}$ such that for every orthonormal basis $\{ \vert h_i 
\rangle \}$ of $\mathfrak{H}$ \[ \sum_i f \left( \vert h_i \rangle 
\right) = \omega \] is satisfied. The constant $\omega$ is called 
the \textsc{weight} of 
the frame function $f$. A frame function $f$ on $\mathfrak{H}$ is 
called \textsc{bounded} if $\sup \left\{ \vert f( \vert h \rangle) 
\vert : \vert h \rangle \in \mathcal{S}(\mathfrak{H}) 
\right\} < \infty$. A frame function $f$ on 
$\mathfrak{H}$ is called \textsc{regular} if there is a Hermitean 
operator $T_f$ on $\mathfrak{H}$ such that $f$ can be written as 
$f \left( \vert h \rangle \right) = \langle h \vert T_f \vert h 
\rangle$, for all $\vert h \rangle \in \mathcal{S}(\mathfrak{H})$, 
where $\langle \cdot \vert 
\cdot \rangle$ denotes the inner product in $\mathfrak{H}$. \\ 
There is a duality between completely additive signed measures on 
$\mathcal{P}(\mathfrak{H})$ and frame functions on $\mathfrak{H}$:
let $m$ be a completely additive signed measure on 
$\mathcal{P}(\mathfrak{H})$ and denote for every $\vert h \rangle 
\in \mathcal{S}(\mathfrak{H})$ the projection operator onto $\vert 
h \rangle$ by $P_h = \vert h \rangle \langle h \vert$, then $f_m 
\left( \vert h \rangle \right) := m(P_h), \vert h \rangle \in 
\mathcal{S}(\mathfrak{H})$, 
defines a frame function $f_m$ on $\mathfrak{H}$ with weight 
$\omega_m = m(1)$. Conversely, 
let $f$ be a frame function on $\mathfrak{H}$. Let $P \in 
\mathcal{P}(\mathfrak{H})$ and let $\{ P_i 
\}$ be a decomposition of $P$ into mutually orthogonal 
one-dimensional projection operators. Denote by $\vert P_i 
\rangle$ the unit vector in $\mathfrak{H}$ onto which the $i$th 
projector $P_i$ projects, then $m_f(P) := \sum_i f \left( \vert 
P_i \rangle \right)$ defines a completely additive 
signed measure $m_f$ on $\mathfrak{H}$. It is easy to see that 
$f_{m_f} 
= f$. \begin{prop} For any integer $n>2$ let $\mathfrak{H}_n$ be 
an $n$-dimensional complex Hilbert space. Then every 
bounded frame function on $\mathfrak{H}_n$ is regular. \label{P1} 
\end{prop}
\begin{prop} Let $\mathfrak{H}$ be an infinite-dimensional 
complex Hilbert space. Then any frame function on $\mathfrak{H}$ 
is bounded and regular. \label{P2} \end{prop}
For the proof of Propositions \ref{P1} and \ref{P2} we refer the 
reader to \cite{Dvurecenskij93}, Chapter 3. \\ 

Now let $\mathfrak{K}$ be a dense linear subspace of 
$\mathfrak{H}$. Define $\mathcal{S}(\mathfrak{K}) := 
\mathcal{S}(\mathfrak{H}) \cap \mathfrak{K}$. A map 
$g : \mathcal{S}(\mathfrak{K}) \to \mathbb{R}$ is said to be a 
\textsc{frame type function on} $\mathfrak{H}$ if the following 
conditions are satisfied \begin{itemize} \item the family $\{ 
g \left( \vert h_i \rangle \right) \}$ is summable for every 
orthonormal system $\{ \vert h_i \rangle \}$ in $\mathfrak{K}$; 
\item for any finite-dimensional subspace $\mathcal{K}_0$ of 
$\mathfrak{K}$, the restriction $g 
\vert_{\mathcal{S}(\mathcal{K}_0)}$ of $g$ to 
$\mathcal{S}(\mathcal{K}_0)$ is 
a frame function on $\mathcal{K}_0$. \end{itemize} Now we have the 
following important result due to Dorofeev and Sherstnev 
\begin{prop} Let $\mathfrak{K}$ be a dense linear subspace of an 
infinite-dimensional complex Hilbert space $\mathfrak{H}$ and let 
$g : \mathcal{S}(\mathfrak{K}) \to \mathbb{R}$ be a frame type 
function on $\mathfrak{H}$. Then $g$ is bounded and there is a 
unique Hermitean trace class operator $T_g$ on $\mathfrak{H}$ 
such that $g( \vert h \rangle) = \langle h \vert T_g \vert h 
\rangle$ for all $\vert h \rangle 
\in \mathcal{S}(\mathfrak{K})$. \label{P3} \end{prop} 
A proof of this proposition can be found in \cite{Dvurecenskij93}, 
Section 3.2.4. 
\subsection{The finite-dimensional case}
\begin{theo} If the dimension $\dim(\mathfrak{H})$ of a 
complex Hilbert space $\mathfrak{H}$ satisfies $2< 
\dim(\mathfrak{H}) < \infty$, then \label{T1} 
there is a one-one correspondence between bounded decoherence 
functionals $d$ on $\mathcal{P}({\mathfrak H})$ and 
trace class operators $\mathfrak{X}$ on ${\mathfrak H} \otimes 
{\mathfrak H}$ according to the rule \begin{equation}
d(h,k) = \mathrm{tr}_{\mathfrak{H} \otimes \mathfrak{H}}(h \otimes 
k \mathfrak{X}) \end{equation}
with the restrictions that \begin{itemize} \item 
$\mathrm{tr}_{\mathfrak{H} \otimes \mathfrak{H}}(h \otimes k 
\mathfrak{X}) = \mathrm{tr}_{\mathfrak{H} \otimes \mathfrak{H}}(k 
\otimes h \mathfrak{X}^{\dagger})$ for all $h,k \in 
\mathcal{P}(\mathfrak{H})$; \item $\mathrm{tr}_{\mathfrak{H} 
\otimes \mathfrak{H}}(h \otimes h \mathfrak{X}) \geq 0$ for all $h 
\in \mathcal{P}(\mathfrak{H})$; \item $\mathrm{tr}_{\mathfrak{H} 
\otimes \mathfrak{H}}(\mathfrak{X}) =1$. \end{itemize} 
In particular, every such decoherence functional is uniformly 
bi-continuous. \end{theo} Theorem \ref{T1} has first been proven 
by Isham, Linden and Schreckenberg in \cite{IshamLS94}. Theorem
\ref{T1} is {\bf not} valid if $\dim(\mathfrak{H}) =2$. \\ 
\textbf{Proof:} Consider the finite-dimensional Hilbert space 
$\mathfrak{H} \otimes \mathfrak{H}$. According to Proposition 
\ref{L6} every bounded decoherence 
functional $d$ on $\mathfrak{H}$ can be extended to a bounded 
functional $\widehat{d} : \mathcal{P}(\mathfrak{H} \otimes 
\mathfrak{H}) \to \mathbb{C}$. Hence, the real part $\Re 
\widehat{d}$ of $\widehat{d}$ induces a bounded frame function 
$f_{\Re d}$ on $\mathfrak{H} \otimes \mathfrak{H}$ by \[ f_{\Re 
d} \left( \vert \Psi_0 \rangle \right) := \Re 
\widehat{d}(P_{\Psi_0}), \] for 
all $\vert \Psi_0 \rangle \in \mathcal{S}(\mathfrak{H} \otimes 
\mathfrak{H})$. Similarly, the imaginary part $\Im \widehat{d}$ of 
$\widehat{d}$ induces a bounded 
frame function $f_{\Im d}$ on $\mathfrak{H} \otimes \mathfrak{H}$ 
by $f_{\Im d} \left( \vert \Psi_0 \rangle \right) := \Im 
\widehat{d}(P_{\Psi_0}),$ for all $\vert \Psi_0 \rangle \in 
\mathcal{S}(\mathfrak{H} \otimes \mathfrak{H})$. 
Therefore, according to Proposition \ref{P1} 
both $f_{\Re d}$ and $f_{\Im d}$ are regular. This proves the 
existence of two Hermitean operators $\mathfrak{X}_{\Re d}$ and 
$\mathfrak{X}_{\Im d}$ on $\mathfrak{H} \otimes \mathfrak{H}$ such 
that $\widehat{d}$ can be written as 
$\widehat{d}(P_M) = \mathrm{tr}_{\mathfrak{H} \otimes 
\mathfrak{H}}(P_M (\mathfrak{X}_{\Re d} + i \mathfrak{X}_{\Im 
d}))$, for all $P_M \in \mathcal{P}_{fin}(\mathfrak{H} \otimes 
\mathfrak{H})$. In particular, it follows $d(h,k) = 
\mathrm{tr}_{\mathfrak{H} \otimes \mathfrak{H}}(h \otimes k 
\mathfrak{X}_d)$ for all $h,k \in \mathcal{P}(\mathfrak{H})$ where 
we have set $\mathfrak{X}_d := \mathfrak{X}_{\Re d} + i 
\mathfrak{X}_{\Im d}$. The remaining assertions of the Theorem are 
now straightforward. \hfill $\blacksquare$ \\ \\ 
Theorem 1 shows that in the finite-dimensional case a decoherence 
functional $d$ on $\mathcal{P}(\mathfrak{H})$ is ultraweakly 
bi-continuous if and only if $d$ is bi-continuous with respect to 
the uniform (or operator norm) topology on 
$\mathcal{P}(\mathfrak{H})$. Moreover, since in the 
finite-dimensional case the weak and the ultraweak topology on 
$\mathcal{P}(\mathfrak{H})$ coincide, an ultraweakly 
bi-continuous decoherence functional $d$ is also weakly 
bi-continuous. 

\subsection{The infinite-dimensional case}
\begin{theo} Let $\mathfrak{H}$ be a complex Hilbert space with 
dimension greater than two, $\dim(\mathfrak{H}) > 2$, then 
\label{T2} there is a one-one correspondence between normal 
completely additive decoherence 
functionals $d$ on $\mathcal{P}({\mathfrak H})$ and 
trace class operators $\mathfrak{X}$ on ${\mathfrak H} \otimes 
{\mathfrak H}$ according to the rule \begin{equation}
d(h,k) = \mathrm{tr}_{\mathfrak{H} \otimes \mathfrak{H}}(h \otimes 
k \mathfrak{X}) \end{equation}
with the restrictions that \begin{itemize} \item 
$\mathrm{tr}_{\mathfrak{H} \otimes \mathfrak{H}}(h \otimes k 
\mathfrak{X}) = \mathrm{tr}_{\mathfrak{H} \otimes \mathfrak{H}}(k 
\otimes h \mathfrak{X}^{\dagger})$ for all $h,k \in 
\mathcal{P}(\mathfrak{H})$; \item $\mathrm{tr}_{\mathfrak{H} 
\otimes \mathfrak{H}}(h \otimes h \mathfrak{X}) \geq 0$ for all $h 
\in \mathcal{P}(\mathfrak{H})$; \item $\mathrm{tr}_{\mathfrak{H} 
\otimes \mathfrak{H}}(\mathfrak{X}) =1$. \end{itemize} \end{theo} 
\textbf{Proof}: If $\mathfrak{H}$ is finite-dimensional, then the 
assertion of the theorem has already been proven in Theorem 
\ref{T1}. Let $\mathfrak{H}$ be infinite-dimensional and let $d$ 
denote a normal decoherence functional 
on $\mathcal{P}(\mathfrak{H})$. Notice, that the requirement of 
complete additivity in the theorem is redundant. As above we 
denote by $\mathfrak{H}_0$ the everywhere dense linear subspace of 
$\mathfrak{H} \otimes \mathfrak{H}$ generated by the simple 
vectors of the form $\vert \phi \rangle \otimes \vert \psi 
\rangle$. From Proposition \ref{L6} we know that $d$ can be 
uniquely extended to a functional $\widehat{d} : 
\mathcal{P}_{fin}(\mathfrak{H}_0) \to \mathbb{C}$. The real 
part $\Re \widehat{d}$ of $\widehat{d}$ induces a frame type 
function $g_{\Re d} : \mathcal{S}(\mathfrak{H}_0) \to \mathbb{R}$ 
on $\mathfrak{H} \otimes \mathfrak{H}$ by \[ g_{\Re 
d} \left( \vert \Psi_0 \rangle \right) := \Re 
\widehat{d}(P_{\Psi_0}), \] for 
all $\vert \Psi_0 \rangle \in \mathcal{S}(\mathfrak{H}_0)$. 
Similarly, the imaginary part $\Im \widehat{d}$ of $\widehat{d}$ 
induces a frame type function 
$g_{\Im d} : \mathcal{S}(\mathfrak{H}_0) \to \mathbb{R}$ on 
$\mathfrak{H} \otimes \mathfrak{H}$ by $g_{\Im d} \left( \vert 
\Psi_0 \rangle \right) := \Im \widehat{d}(P_{\Psi_0}),$ 
for all $\vert \Psi_0 \rangle \in \mathcal{S}(\mathfrak{H}_0)$. 
According to Proposition \ref{P3} both $g_{\Re d}$ and $g_{\Im d}$ 
are regular. This proves the 
existence of two Hermitean operators $\mathfrak{X}_{\Re d}$ and 
$\mathfrak{X}_{\Im d}$ on $\mathfrak{H} \otimes \mathfrak{H}$ such 
that $\widehat{d}$ can be written as 
$\widehat{d}(P_M) = \mathrm{tr}_{\mathfrak{H} \otimes 
\mathfrak{H}}(P_M (\mathfrak{X}_{\Re d} + i \mathfrak{X}_{\Im 
d})) = \mathrm{tr}_{\mathfrak{H} \otimes 
\mathfrak{H}}(P_M \mathfrak{X}_d)$, for all $P_M \in 
\mathcal{P}_{fin}(\mathfrak{H}_0)$ where we have set 
$\mathfrak{X}_d := \mathfrak{X}_{\Re d} + i 
\mathfrak{X}_{\Im d}$. It is clear now, that the real part 
$\Re \widehat{d}$ of $\widehat{d}$ can 
be extended to a Jordan completely additive 
signed measure $\Re \bar{d}$ on $\mathfrak{H} \otimes 
\mathfrak{H}$ given by $\Re \bar{d}(P) = \mathrm{tr}_{\mathfrak{H} 
\otimes 
\mathfrak{H}}(P \mathfrak{X}_{\Re d})$ for all $P \in 
\mathcal{P}(\mathfrak{H} \otimes \mathfrak{H})$. The imaginary 
part $\Im \widehat{d}$ of $\widehat{d}$ can be extended to a 
Jordan completely additive 
signed measure $\Im \bar{d}$ on $\mathfrak{H} \otimes 
\mathfrak{H}$ given by $\Im \bar{d}(P) = 
\mathrm{tr}_{\mathfrak{H} \otimes \mathfrak{H}}(P 
\mathfrak{X}_{\Im d})$ for all $P \in 
\mathcal{P}(\mathfrak{H} \otimes \mathfrak{H})$. Hence 
$\widehat{d}$ can be extended to a 
completely additive complex valued measure $\bar{d}$ on 
$\mathfrak{H} \otimes \mathfrak{H}$ given by \[ \bar{d}(P) = 
\mathrm{tr}_{\mathfrak{H} \otimes \mathfrak{H}}(P \mathfrak{X}_d), 
\] for all $P \in \mathcal{P}(\mathfrak{H} \otimes \mathfrak{H})$.
Since $d$ is finitely additive, it follows that $d(h,k) = 
\bar{d}(h \otimes k)$ for all finite-dimensional $h,k \in 
\mathcal{P}(\mathfrak{H})$. Obviously $\bar{d}$ is 
ultraweakly continuous. By assumption $d$ is completely additive 
in both arguments. Now let $h,k \in 
\mathcal{P}(\mathfrak{H})$ denote two arbitrary projection 
operators on $\mathfrak{H}$. Consider the projection operator $h 
\otimes k$ on $\mathfrak{H} \otimes \mathfrak{H}$. Then there is a 
family of mutually orthogonal one-dimensional projection operators 
$\{ h_i \}_{i \in I}$ such that $h = \sum_i h_i$ in the weak 
operator topology and a family of mutually orthogonal 
one-dimensional projection operators $\{ k_l \}_{l \in L}$ such 
that $k = \sum_l k_l$ in the weak operator topology, compare, 
e.g., \cite{Naimark70} Section I.5.12. Obviously $h 
\otimes k = \sum_{i,l} h_i \otimes k_l$ in the weak operator 
topology. Consider the net $\mathcal{N}$ consisting of all finite 
sums of the form $\left( \sum_{i \in \widetilde{I}} 
h_i \right) \otimes \left( \sum_{l \in \widetilde{L}} k_l \right)$ 
where $\widetilde{I}$ is a finite subset of $I$ and 
$\widetilde{L}$ is a finite subset of $L$. 
Proposition I.5.12.IX in \cite{Naimark70} implies that for 
every trace class operator $T$ on $\mathfrak{H} \otimes 
\mathfrak{H}$ there exist at most countable subsets $I_T \subset 
I$ and $L_T \subset L$ such that $\mathrm{tr}_{\mathfrak{H} 
\otimes 
\mathfrak{H}}(h \otimes k T) = \mathrm{tr}_{\mathfrak{H} \otimes 
\mathfrak{H}} \left( \left( \sum_{i \in I_T} h_i \right) \otimes 
\left( \sum_{l \in L_T} k_l \right) T \right)$. Hence, for all 
$\epsilon > 0$ and all trace class 
operators $T$ on $\mathfrak{H} \otimes \mathfrak{H}$ there exist 
finite subsets $\widetilde{I}_{T, \epsilon}$ of $I$ and 
$\widetilde{L}_{T, \epsilon}$ of 
$L$ such that $\left\vert \mathrm{tr}_{\mathfrak{H} \otimes 
\mathfrak{H}}(h \otimes k T) - \mathrm{tr}_{\mathfrak{H} \otimes 
\mathfrak{H}} \left( \left( \sum_{i \in \widetilde{I}} h_i \right) 
\otimes \left( \sum_{l \in \widetilde{L}} k_l \right) T \right) 
\right\vert < \epsilon$ for all 
finite $\widetilde{I} \supset \widetilde{I}_{T, \epsilon}$ and all 
finite $\widetilde{L} \supset \widetilde{L}_{T, \epsilon}$. We 
conclude that the net $\mathcal{N}$ defined above converges to $h 
\otimes k$ in the ultraweak topology. Clearly, the net of 
complex numbers $ \left\{ \bar{d} \left( \left( \sum_{i \in 
\widetilde{I}} h_i \right) 
\otimes \left( \sum_{l \in \widetilde{L}} k_l \right) \right) 
\right\}_{\widetilde{I}, 
\widetilde{L}}$, where $\widetilde{I}$ and $\widetilde{L}$ run 
through all finite subsets of $I$ and $L$ respectively, converges 
to $\bar{d}(h \otimes k)$. Similarly, the 
net $\left\{ d \left( \sum_{i \in \widetilde{I}} h_i, \sum_{l 
\in \widetilde{L}} k_l \right) \right\}_{\widetilde{I}, 
\widetilde{L}}$, where $\widetilde{I}$ and $\widetilde{L}$ run 
through all finite subsets of $I$ and $L$ respectively, 
converges to $d(h,k)$. Since $\bar{d} \left( \left( \sum_{i \in 
\widetilde{I}} h_i \right) \otimes \left( \sum_{l \in 
\widetilde{L}} k_l \right) \right) =  d \left( \sum_{i \in 
\widetilde{I}} h_i, \sum_{l \in \widetilde{L}} k_l \right)$ 
for every finite $\widetilde{I} \subset I$ and $\widetilde{L} 
\subset L$, we conclude that 
$\bar{d}(h \otimes k) = d(h,k)$. In particular, it follows $d(h,k) 
= \mathrm{tr}_{\mathfrak{H} \otimes \mathfrak{H}}(h \otimes k 
\mathfrak{X}_d)$ for all $h,k \in \mathcal{P}(\mathfrak{H})$. The 
remaining assertions of the Theorem are 
now straightforward. \hfill $\blacksquare$ \\ \\
We say that a decoherence functional $d$ on the set 
$\mathcal{P}(\mathfrak{H})$ of projectors on the Hilbert space 
$\mathfrak{H}$ is \textsc{regular} if there is a trace class 
operator $\mathfrak{X}_d$ on $\mathfrak{H} \otimes \mathfrak{H}$ 
such that $d$ can be written as $d(h,k) = 
\mathrm{tr}_{\mathfrak{H} \otimes \mathfrak{H}}(h \otimes k 
\mathfrak{X}_d)$ for all $h,k \in \mathcal{P}(\mathfrak{H})$. In 
this case we say that $\mathfrak{X}_d$ defines a \textsc{regular 
representation} of $d$. \\
Moreover, we say that a decoherence functional $d$ on the set 
$\mathcal{P}(\mathfrak{H})$ of projectors on the Hilbert space 
$\mathfrak{H}$ is \textsc{quasi-regular} if there is a bounded 
operator $\mathfrak{X}_d$ on $\mathfrak{H} \otimes \mathfrak{H}$ 
(not necessarily of trace class) such that $d$ can be written as 
$d(h,k) = \mathrm{tr}_{\mathfrak{H} \otimes \mathfrak{H}}(h 
\otimes k \mathfrak{X}_d)$ for all finite-dimensional $h,k \in 
\mathcal{P}(\mathfrak{H})$. In this case we say that 
$\mathfrak{X}_d$ defines a \textsc{quasi-regular representation} 
of $d$. \\
We say that a decoherence functional $d$ on the set 
$\mathcal{P}(\mathfrak{H})$ of projectors on the Hilbert space 
$\mathfrak{H}$ is 
$\sigma$-\textsc{quasi-regular} if there is a bounded 
operator $\mathfrak{X}_d$ on $\mathfrak{H} \otimes \mathfrak{H}$ 
(not necessarily of trace class) such that the
following condition is satisfied: given any two projection 
operators $h,k \in \mathcal{P}(\mathfrak{H})$ projecting onto 
separable subspaces of $\mathfrak{H}$ and arbitrary decompositions 
$\{ h_i \}$ of $h$ and $\{k_j \}$ of $k$ into mutually orthogonal 
one-dimensional projection operators $h = \sum_i h_i$ and $k = 
\sum_j k_j$, then the sum $\sum_{i,j} \mathrm{tr}_{\mathfrak{H} 
\otimes \mathfrak{H}}(h_i \otimes k_j \mathfrak{X}_d)$ is 
well-defined and independent of the particular decompositions 
considered and equals $d(h,k)$, i.e., $d(h,k) = \sum_{i,j} 
\mathrm{tr}_{\mathfrak{H} \otimes \mathfrak{H}}(h_i \otimes k_j 
\mathfrak{X}_d) = \sum_{i,j} \langle h_i \otimes k_j \vert 
\mathfrak{X}_d \vert h_i \otimes k_j \rangle$. In this case we say 
that $\mathfrak{X}_d$ defines a $\sigma$-\textsc{quasi-regular 
representation} of $d$. \\ 
We say that a decoherence functional $d$ on the set 
$\mathcal{P}(\mathfrak{H})$ of projectors on the Hilbert space 
$\mathfrak{H}$ is \textsc{pseudo-regular} if there is a bounded 
operator $\mathfrak{X}_d$ on $\mathfrak{H} \otimes \mathfrak{H}$ 
(not necessarily of trace class) such that the
following condition is satisfied: given any two projection 
operators $h,k \in \mathcal{P}(\mathfrak{H})$ projecting onto 
arbitrary subspaces of $\mathfrak{H}$ and given arbitrary 
decompositions 
$\{ h_i \}$ of $h$ and $\{k_j \}$ of $k$ into mutually orthogonal 
one-dimensional projection operators $h = \sum_i h_i$ and $k = 
\sum_j k_j$, then the family $\{ \sum_{i,j} 
\mathrm{tr}_{\mathfrak{H} \otimes \mathfrak{H}}(h_i \otimes k_j 
\mathfrak{X}_d) \}$ is summable and its sum is independent of the 
particular decompositions 
considered and equals $d(h,k)$, i.e., $d(h,k) = \sum_{i,j} 
\mathrm{tr}_{\mathfrak{H} \otimes \mathfrak{H}}(h_i \otimes k_j 
\mathfrak{X}_d) = \sum_{i,j} \langle h_i \otimes k_j \vert 
\mathfrak{X}_d \vert h_i \otimes k_j \rangle$. In this case we say 
that $\mathfrak{X}_d$ defines a \textsc{pseudo-regular 
representation} of $d$. 
\begin{co} Let $\mathfrak{H}$ be an infinite-dimensional complex 
Hilbert space. Then every completely additive normal decoherence 
functional $d$ on $\mathcal{P}(\mathfrak{H})$ is ultraweakly 
bi-continuous and regular. \end{co}
In the following propositions the requirement that the decoherence 
functional is normal is weakened. 
\begin{prop} Let $\mathfrak{H}$ be an infinitely dimensional 
complex Hilbert space, then \label{T7} for every proper completely 
additive decoherence functional $d$ on $\mathcal{P}({\mathfrak 
H})$ there exists a unique Hilbert-Schmidt operator 
$\mathfrak{X}_d$ on $\mathfrak{H} \otimes \mathfrak{H}$ 
(not necessarily of trace class) defining a pseudo-regular 
representation of $d$. \end{prop} 
\textbf{Proof:} Denote by $\widehat{d}$ the extension of $d$ from 
Proposition \ref{L6}. Since $\widehat{d}$ is proper, it follows by 
a standard argument (see, 
e.g., the first part of the proof of Theorem 3.2.21 in 
\cite{Dvurecenskij93}) that there exist uniquely determined 
bounded Hermitean operators $\mathfrak{X}_{\Re d}$ and 
$\mathfrak{X}_{\Im d}$ on $\mathfrak{H} \otimes 
\mathfrak{H}$ such that $\Re \widehat{d}$ and $\Im \widehat{d}$ 
can be written as $\Re \widehat{d} \left( P_{\Psi_0} \right) = 
\langle \Psi_0 \vert 
\mathfrak{X}_{\Re d} \vert \Psi_0 \rangle$ and $\Im \widehat{d} 
\left( P_{\Psi_0} \right) = \langle \Psi_0 \vert 
\mathfrak{X}_{\Im d} \vert \Psi_0 \rangle$ for all $\vert \Psi_0 
\rangle \in \mathcal{S}(\mathfrak{H}_0)$. 
Denote by $\{ \varphi_i \}$ and $\{ \chi_j \}$ two complete 
systems of mutually orthogonal one-dimensional projection 
operators on $\mathfrak{H}$ and by $\{ \vert \varphi_i \rangle \} 
$ and $\{ \vert \chi_j \rangle \}$ the corresponding orthonormal 
bases. Since $d$ is completely additive, it follows $\sum_{i,j} 
\vert \langle \varphi_i \otimes \chi_j \vert \mathfrak{X}_{\Re d} 
\vert \varphi_i \otimes \chi_j \rangle \vert^2 = \sum \vert \Re 
d(\varphi_i, \chi_j) \vert^2 \leq  \left(\sum \vert \Re 
d(\varphi_i, \chi_j) \vert \right)^2 < \infty.$ Hence 
$\mathfrak{X}_{\Re d}$ is a Hilbert-Schmidt operator. Similarly, 
$\mathfrak{X}_{\Im d}$ is a Hilbert-Schmidt operator. Then define 
$\mathfrak{X}_d \equiv \mathfrak{X}_{\Re d} + i \mathfrak{X}_{\Im 
d}$. \hfill $\blacksquare$ 
\begin{prop} Let $\mathfrak{H}$ be an infinitely dimensional 
complex Hilbert space, then \label{T5} for every proper finitely 
additive decoherence functional $d$ on $\mathcal{P}({\mathfrak 
H})$ there exists a unique bounded operator $\mathfrak{X}_d$ on 
$\mathfrak{H} \otimes \mathfrak{H}$ 
(not necessarily of trace class) such that $d$ can be written as 
$d(h,k) = \mathrm{tr}_{\mathfrak{H} \otimes \mathfrak{H}}(h 
\otimes k \mathfrak{X}_d)$ for all finite-dimensional $h,k \in 
\mathcal{P}(\mathfrak{H})$. \end{prop} 
In other words: every proper decoherence functional admits a 
\emph{unique} quasi-regular representation. 
\begin{prop} Let $\mathfrak{H}$ be an infinitely dimensional 
complex Hilbert space, then \label{T4} for every proper 
$\sigma$-additive decoherence functional $d$ on 
$\mathcal{P}({\mathfrak H})$ there exists a unique bounded 
operator $\mathfrak{X}_d$ on $\mathfrak{H} \otimes \mathfrak{H}$ 
(not necessarily of trace class) defining a $\sigma$-quasi-regular 
representation of $d$. If $\mathfrak{H}$ is separable and if $d$ 
is $\sigma$-summable, then 
$\mathfrak{X}_d$ is a Hilbert-Schmidt operator. \end{prop} 
The proofs of Proposition \ref{T5} and Proposition \ref{T4} are 
analoguous to 
the proof of Theorem \ref{T2} and Proposition \ref{T7}. 
\begin{co} Let $\mathfrak{H}$ be an infinite-dimensional complex 
separable Hilbert space. Then every $\sigma$-summable normal 
decoherence functional $d$ on $\mathcal{P}(\mathfrak{H})$ is 
ultraweakly bi-continuous and regular. \end{co}
To sum up: Theorem \ref{T1} asserts that every bounded decoherence 
functional $d$ on the set of projection operators on a 
finite-dimensional Hilbert space of dimension 
greater than two is regular. Theorem \ref{T2} shows that every 
normal completely additive decoherence functional $d$ on 
the set of projection operators of an 
infinite-dimensional Hilbert space is regular. And Proposition 
\ref{T4} shows among others that every $\sigma$-summable normal 
decoherence 
functional $d$ on the set of projection operators of an 
infinite-dimensional separable Hilbert space is regular. Obviously 
every regular decoherence functional is ultraweakly bi-continuous 
and hence $\sigma$-summable for separable Hilbert spaces and 
completely additive for non-separable Hilbert spaces. 
\subsection{Effect Histories}
In \cite{Rudolph96} it has been argued that in a general history 
theory the space of histories should be identified with the set of 
effects $\mathfrak{E}(\mathfrak{H})$ on some Hilbert space 
$\mathfrak{H}$. The following Corollary \ref{L1} shows that there 
is a one-one correspondence between ultraweakly continuous 
normal decoherence functionals $d : \mathcal{P}(\mathfrak{H}) 
\times \mathcal{P}(\mathfrak{H}) \to \mathbb{C}$ and ultraweakly 
continuous normal decoherence functionals $d : 
\mathfrak{E}(\mathfrak{H}) \times \mathfrak{E}(\mathfrak{H}) \to 
\mathbb{C}$ as defined in \cite{Rudolph96}. Corollary \ref{L1} is 
an easy consequence of Theorem \ref{T2}. 
\begin{co} Let $d$ denote a completely additive normal 
decoherence functional on $\mathcal{P}({\mathfrak H})$, 
$\dim(\mathfrak{H})$ $ > 2$, then $d$ 
can be uniquely extended to an ultraweakly bi-continuous bounded 
functional $\widetilde{d} : \mathfrak{E}(\mathfrak{H}) \times 
\mathfrak{E}(\mathfrak{H}) \to \mathbb{C}$. $\widetilde{d}$ 
satisfies $\widetilde{d}(e,e) \in \mathbb{R}; 
\widetilde{d}(e,e)\geq 0; \widetilde{d}(e,f) = 
\widetilde{d}(f,e)^*; 
\widetilde{d}(1,1)=1$ and $\widetilde{d}(0,e)=0$, for all $e,f \in 
\mathfrak{E}(\mathfrak{H})$. Moreover, $\widetilde{d}$ is additive 
with respect to the canonical D-poset structure on 
$\mathfrak{E}(\mathfrak{H})$, i.e., $\widetilde{d}(e_1 \oplus e_2, 
f) = \widetilde{d}(e_1,f) + \widetilde{d}(e_2,f)$, whenever $e_1 
\oplus e_2$ is well-defined. \label{L1} \end{co} 

\section{Summary}
In this work we have proven a classification theorem for 
decoherence functionals on the set $\mathcal{P}(\mathfrak{H})$ of 
projection operators on an arbitrary finite- or 
infinite-dimensional separable or non-separable complex Hilbert 
space $\mathfrak{H}$ with dimension greater than two. In the 
finite-dimensional case we have seen that there is a one-to-one 
correspondence between bounded decoherence functionals on 
$\mathcal{P}(\mathfrak{H})$ and certain trace class operators 
$\mathfrak{X}$ on $\mathfrak{H} \otimes \mathfrak{H}$. The 
conditions $\mathfrak{X}$ has to satisfy are listed in Theorem 
\ref{T1}. This result has first been proven by Isham, Linden 
and Schreckenberg \cite{IshamLS94}. 
If $\mathfrak{H}$ is an infinite-dimensional 
separable Hilbert space, we have shown that there is a one-to-one 
correspondence between normal ($\sigma$-summable) decoherence 
functionals on $\mathcal{P}(\mathfrak{H})$ and certain trace class 
operators $\mathfrak{X}$ on $\mathfrak{H} \otimes \mathfrak{H}$. 
The conditions $\mathfrak{X}$ has to satisfy are listed in Theorem 
\ref{T2}. If $\mathfrak{H}$ is an 
infinite-dimensional non-separable Hilbert space, we have seen 
that there is a one-to-one correspondence between normal 
(completely additive) decoherence functionals on 
$\mathcal{P}(\mathfrak{H})$ and certain trace class operators 
$\mathfrak{X}$ on $\mathfrak{H} \otimes \mathfrak{H}$. The 
conditions $\mathfrak{X}$ has to satisfy are listed in Theorem 
\ref{T2}. \\ In addition, we have seen that if $\mathfrak{H}$ 
is an arbitrary Hilbert space with $\dim(\mathfrak{H}) > 2$, then 
every proper decoherence functional on $\mathcal{P}(\mathfrak{H})$ 
admits a unique quasi-regular representation, every proper 
$\sigma$-additive 
decoherence functional admits a unique $\sigma$-quasi-regular 
representation and every proper completely additive 
decoherence functional admits a unique pseudo-regular 
representation. \\  
There are many decoherence functionals which are not covered by 
Theorem \ref{T2} and the subsequent propositions. It would be 
particularly interesting to learn more 
about the general structure and properties of the quasi-regular 
representations of decoherence functionals and about 
representations for non-normal 
and non-proper decoherence functionals. These topics deserve 
further investigation. 

\subsubsection*{Acknowledgments}
I am grateful to Professor Frank Steiner for his 
support of my work. I would like to thank Dr.~Stephan 
Schreckenberg for stimulating discussions. Financial support given 
by Deutsche Forschungsgemeinschaft (Graduiertenkolleg f\"ur 
theoretische Elementarteilchenphysik) is also gratefully 
acknowledged. I am grateful to Professor J.D.M.~Wright for his 
insightful comments on a previous version of this paper.

\end{document}